# Sailing the Information Ocean with Awareness of Currents: Discovery and Application of Source Dependence

Laure Berti-Equille*
Université de Rennes 1
berti@irisa.fr

Anish Das Sarma
Stanford University
anish@cs.stanford.edu

Xin Luna Dong
AT&T Labs–Research
lunadong@research.att.com

Amélie Marian
Rutgers University
amelie@cs.rutgers.edu

Divesh Srivastava
AT&T Labs–Research
divesh@research.att.com

## ABSTRACT

The Web has enabled the availability of a huge amount of useful information, but has also eased the ability to spread false information and rumors across multiple sources, making it hard to distinguish between what is true and what is not. Recent examples include the premature Steve Jobs obituary, the second bankruptcy of United airlines, the creation of Black Holes by the operation of the Large Hadron Collider, etc. Since it is important to permit the expression of dissenting and conflicting opinions, it would be a fallacy to try to ensure that the Web provides only consistent information. However, to help in separating the wheat from the chaff, it is essential to be able to determine dependence between sources. Given the huge number of data sources and the vast volume of conflicting data available on the Web, doing so in a scalable manner is extremely challenging and has not been addressed by existing work yet.

In this paper, we present a set of research problems and propose some preliminary solutions on the issues involved in discovering dependence between sources. We also discuss how this knowledge can benefit a variety of technologies, such as data integration and Web 2.0, that help users manage and access the totality of the available information from various sources.

## 1. INTRODUCTION

The amount of useful information available on the Web has been growing at a dramatic pace in recent years. In a variety of domains, such as science, business, technology, arts, entertainment, politics, government, sports, travel, there are a huge number of data sources that seek to provide information to a wide spectrum of information users. In addition to enabling the availability of useful information, the Web has also eased the ability to publish and spread false information across multiple sources. For example, an obituary of Apple founder Steve Jobs was published and sent to thousands of corporate clients on Aug 28, 2008, before it was retracted.[1] Such false information can often result in considerable damage; for example, the recent incorrect news about United airlines filing for a second bankruptcy sent its shares tumbling, before the error was corrected.[2] The Web also makes it easy to rapidly spread rumors, which take a long time to die down. For example, the rumor from the late 1990s that the MMR vaccine given to children in Britain was harmful and linked to autism caused a significant drop in MMR coverage, leading autism experts to spend years trying to dispel the rumor.[3] Similarly, the upcoming experiments at the Large Hadron Collider (LHC) have sparked fears among the public that the LHC particle collisions might produce dangerous microscopic black holes that may mean the end of the world.[4]

Widespread availability of conflicting information (some true, some false) makes it hard to distinguish between what is true and what is not. However, it would be a fallacy to try to ensure, even if it were feasible, that the Web provides only consistent information. As the following quote[5], often mis-attributed to Voltaire, eloquently expressed, it is important in a free society to permit the expression of dissenting and conflicting opinions.

> I disapprove of what you say, but I will defend to the death your right to say it.
> – Evelyn Beatrice Hall

In the presence of conflicting data sources, it is not easy to separate the wheat from the chaff. Simply using the information that is asserted by the largest number of data sources is clearly inadequate since biased (and even malicious) sources abound, and plagiarism (*i.e.*, copying without proper attribution) between sources may be widespread. How can one find good answers to queries in such a "bad world"? We argue that, in order to be able to do so, *it is essential to be able to determine dependence between data sources*. Dependence between sources can arise when a source copies values from another source, or when a source chooses to provide values that conflict with those provided by another source. We refer to the former as *similarity-dependence* and the latter as *dissimilarity-dependence*; both can be readily observed in contentious domains like politics. Given the huge number of data sources and the vast volume of conflicting data available on the Web, determining dependence between sources in a scalable manner is extremely challenging and has not been addressed by existing work yet.

---

*Currently visiting researcher at AT&T Labs–Research, supported by a Marie Curie International Fellowship within the 6th European Community Framework Programme (FP6-MOIF-CT-2006-041000).

[1] http://www.telegraph.co.uk/news/newstopics/howaboutthat/2638481/Steve-Jobs-obituary-published-by-Bloomberg.html

[2] http://gawker.com/5047763/how-robots-destroyed-united-airlines

[3] http://www.guardian.co.uk/society/2008/apr/12/health.children

[4] http://en.wikipedia.org/wiki/Large_Hadron_Collider#Safety_of_particle_collisions

[5] http://en.wikipedia.org/wiki/Evelyn_Beatrice_Hall

Knowledge of dependence between sources can help a variety of technologies, such as information retrieval, news aggregation, web search, data integration, and Web 2.0, that have emerged over the years to help users cope with the explosion of information available from various sources. Here we illustrate with two applications that manage structured information from various sources.

- Data integration systems [11, 12] present a unified way to model and access a diversity of data sources in a given domain. However, query answering in such systems typically assume that the information provided by the sources is consistent; that is, even when their representations in different sources differ because of different schemas or naming conventions, the data sources do not provide conflicting information. Even when the sources contain probabilistic data to represent conflicts, the data integration systems usually combine the probabilities by assuming that the sources are independent.

  Knowing the dependence between sources can help data integration systems identify which value (among a set of conflicting values) should be presented to users.

- Web 2.0 [2] has enabled users to provide information, opinions, and ratings on a wide variety of topics. Technologies such as recommendation systems [1] make use of this user-provided information to aggregate ratings (of products, movies, etc.) from a large set of raters to present users with information targeted to their interests. While recommendation systems are naturally designed to permit conflicting ratings (some raters might like a product or a movie, while others may hate it), they usually assume that the raters are independent of each other; this permits them to uniformly aggregate the ratings across a large set of raters when making recommendations.

  When users copy information from others, or are influenced by other users in the information they provide, they are no longer independent of each other. Knowing the dependence between them can help recommendation systems determine how to aggregate the ratings across the raters to make better (unbiased) recommendations.

In this paper, we propose a set of research problems that together aim to scalably analyze the (possibly conflicting) information provided by (possibly inaccurate, biased, non-independent) structured data sources to (i) discover potential dependence between the sources, and (ii) leverage the knowledge of such dependence to enable technologies such as data integration systems and recommendation systems to effectively deal with the "bad world" scenario. We describe the challenges we face, propose preliminary solutions, and lay down a research agenda for the Database community.

## 2. WHAT IS SOURCE DEPENDENCE?

### 2.1 Data Sources

A structured data source provides information about a set of entities at specific points in time. For example, online bookstores typically provide bibliographic data (*e.g.*, title, authors, publisher, year) and opinions (*e.g.*, ratings, reviews) for a set of books. Financial sites often provide a variety of information about corporations, such as senior officers and their compensation packages, stock price evolution, and opinions about the financial health of the corporation.

Over time, the information associated with an entity may change, and even the set of entities about which a source provides information may change. For example, online bookstores continually add new opinions about existing books to their websites, add information about new books as they are published, modify information about the availability of books, correct potential errors, etc. In the case of financial sites, the officers of a corporation and their compensation packages naturally change over time, as do opinions about their financial health, and financial sites update information on their sites to reflect the changing nature of reality.

We adopt a model of a structured data source $S_i$ as a set of 4-tuples $\{(id_j, time_j, val_j, prob_j)\}$, where $id_j$ is an identifier associated with the value $val_j$ at time $time_j$, with probability $prob_j$. We do not constrain the possible domain of values $val_j$. In the case of relational databases, if the value $val_j$ is a cell value (an atomic value or a list of atomic values), the identifier $id_j$ can encapsulate the table name, record identifier, and column name associated with that cell. If, on the other hand, the value $val_j$ is a tuple, the identifier $id_j$ may represent only the table name and the record identifier. Not all sources may explicitly specify temporal information, in which case $time_j$ may either be inferred from snapshots or be missing altogether. Similarly, when sources do not explicitly specify probability information, $prob_j$ may be assumed to be 1. Similar considerations apply to other kinds of structured data sources, including object-oriented and XML databases, probabilistic databases, document collections that use Information Extraction techniques, etc.

The values provided by different data sources $S_i, S_k$ for an identifier $id_j$ at time $time_j$ may differ. Some of these differences may be representational differences (*e.g.*, "AT&T Research" versus "AT&T Labs–Research") while others can be genuine conflicts (*e.g.*, "AT&T Labs–Research" versus "Rutgers University"). It is the presence of *conflicting* values between sources, possibly along with representational differences, that interests us in this paper.

Some of these conflicting values may be factual differences (*e.g.*, the title of a book with a particular ISBN, or the stock price of a corporation at the end of a particular day of trading), where there is an underlying true value and one can seek to discover the truth from amongst the conflicting values. Other conflicts may represent differences of opinion (*e.g.*, ratings associated with books or restaurants) with no underlying true value, where one can seek to discover a consensus value or identify meaningful differences of opinion. In this paper, we consider both kinds of conflicting values and presume that we know if there is an underlying truth.

### 2.2 Dependence Between Sources

In the presence of conflicting values provided by multiple data sources, discovering the truth or obtaining an unbiased consensus value is not easy. A straightforward approach such as taking the value asserted by the most number of data sources (*i.e.*, naive voting) can result in erroneous results in the presence of data source plagiarism (*i.e.*, copying without proper attribution), as the following examples illustrate.

EXAMPLE 2.1. *Consider sources $S_1$, $S_2$ and $S_3$ in Table 1. Assume that they independently provide information on affiliations of five researchers and only $S_1$ provides the true values for all five researchers. Using naive voting amongst $S_1, S_2$ and $S_3$, we are able to find the correct affiliations of the first four researchers, but remain unsure of the affiliation of Dong as $S_1$, $S_2$ and $S_3$ each votes for a different affiliation.*

*Now consider two other data sources, $S_4$ and $S_5$, in Table 1.*

**Table 1: Researcher affiliation example. Only $S_1$ provides all true values; $S_4$ and $S_5$ copy.**

|  | $S_1$ | $S_2$ | $S_3$ | $S_4$ | $S_5$ |
|---|---|---|---|---|---|
| Suciu | UW | MSR | UW | UW | UWisc |
| Halevy | Google | Google | UW | UW | UW |
| Balazinska | UW | UW | UW | UW | UW |
| Dalvi | Yahoo! | Yahoo! | UW | UW | UW |
| Dong | AT&T | Google | UW | UW | UW |

**Table 2: Movie rating example. Reviewer $R_4$ always disagrees with what $R_1$ rates.**

|  | $R_1$ | $R_2$ | $R_3$ | $R_4$ |
|---|---|---|---|---|
| *The Pianist* | Good | Neutral | Bad | Bad |
| *Into the Wild* | Good | Bad | Good | Bad |
| *The Matrix* | Bad | Bad | Good | Good |

*Assume that they copy their data from $S_3$: $S_4$ makes an exact copy and $S_5$ makes a change during the copying process. A naive voting using sources $S_1$–$S_5$ would select all the values provided by $S_3$ and so makes wrong decisions for three out of five researchers.* □

Example 2.1 illustrates one kind of dependence between data sources, arising from a source copying, possibly of its own volition, the values provided by another data source. We refer to this kind of dependence as *similarity-dependence*, which has the effect of boosting the vote count of values provided by a source under naive voting. Thus, $S_3$ and $S_4$ are said to be similarity-dependent.

EXAMPLE 2.2. *Consider a scenario with movie reviewers, as shown in Table 2. Whereas reviewers $R_1$, $R_2$ and $R_3$ provide independent ratings, $R_4$ has a strong opinion on $R_1$'s tastes and chooses to provide opposite ratings for all of $R_1$'s ratings. A naive aggregation of ratings from reviewers $R_1$–$R_4$ would significantly differ from the aggregation without considering $R_4$.* □

Example 2.2 illustrates a second kind of dependence between data sources, arising from a source choosing to provide values that conflict with the values provided by another data source. We refer to this kind of dependence between sources as *dissimilarity-dependence*, which has the effect of canceling the vote count of values provided by another source under naive voting. Thus, sources $R_1$ and $R_4$ are said to be dissimilarity-dependent.

One can readily observe the presence of similarity-dependence and dissimilarity-dependence between sources, especially in contentious domains like politics. Identifying these kinds of dependence between sources can be invaluable when discovering truth, identifying unbiased consensus values, or characterizing benevolent versus malicious sources.

## 3. SOURCE DEPENDENCE DISCOVERY

### 3.1 Challenges

The preceding discussion on the nature of similarity-dependence and dissimilarity-dependence between data sources suggests that one should be able to determine dependence by comparing the information provided by the sources. The higher the similarity between the data sources, the more is the likelihood of similarity-dependence. Also, the higher the dissimilarity between the data sources, the more is the likelihood of dissimilarity-dependence. However, in practice we need to address many challenges that make this simplistic approach inadequate. We identify a few such challenges next.

**Accurate sources:** Consider the case of information where there is an underlying true value (*e.g.*, bibliographic data for books, affiliation information for researchers). Accurate sources that independently provide true values would be determined as having a high similarity, which might lead to the erroneous conclusion that they are dependent. Even if the sources are not completely accurate, a few errors may not significantly affect the extent of their similarity. For example, one expects there to be a high similarity between the bibliographic information provided by booksellers such as Barnes&Noble and Borders, even though there may be no dependence between the sources. Hence, any approach for determining dependence needs to consider the possibility that the similarity between sources may be due to the sources independently providing true values.

**Different coverage and expertise:** Not all data sources are likely to provide values for every identifier. In many domains such as biological sciences, some sources are *specialist* sources, providing accurate information for a small subset of identifiers, while other sources are *generalist* sources, providing a huge amount of information, some of which may be out-of-date. In addition, some specialist sources may copy from each other for information in areas outside their expertise, resulting in loop copying. This phenomenon has parallels with the notions of authorities and hubs on the Web. Determining dependence between such sources based only on similarity and dissimilarity can quite easily lead to erroneous conclusions. Hence, any approach for determining dependence needs to consider the possibility that sources have different expertise and provide different amounts of information.

**Lazy copiers and slow providers:** Consider the case where the value associated with an identifier may change over time (*e.g.*, primary affiliations of researchers). When an original data source updates its data over time, its copier may be lazy in updating (*e.g.*, the updated data may be copied only in a batch mode), or it may not copy certain kinds of updates (*e.g.*, deleting affiliations), etc. In such cases, the similarity between sources might again be low, leading to the erroneous conclusion that the sources are likely to be independent. On the other hand, an independent source may be slow and often behind other sources in updating values, and so appears to be a copier. Hence, any approach for determining dependence needs to consider the possibility that a source may be slow in copying or in providing independent information.

**Partial dependence:** Even if a data source copies from another source, it may copy only a subset of the information (*e.g.*, only presidential politics in a political source), reformat some of the copied information (*e.g.*, for consistency with local naming conventions), and provide other information independently (*e.g.*, local politics). Similarly, a data source that chooses to provide conflicting information may do so for only a subset of the available information (*e.g.*, on environmental issues in a political source). In such cases, the similarity (resp., dissimilarity) between the sources might not always be high, leading to the erroneous conclusion that the sources are likely to be independent. Hence, any approach for determining dependence needs to consider the possibility that the dependence between sources is only on a subset of the information.

**Correlated information:** Consider the case of information where there may be no underlying truth value (*e.g.*, ratings for movies, responses to opinion polls). A high similarity between the ratings of two raters for the various Star Wars movies may simply reflect a popular opinion amongst science fiction fans about the Star Wars movies, rather than any copying, but might lead to the erroneous conclusion that they are dependent. Similarly, a high similarity between the opinions of two sources on a narrow set of political questions may simply reflect their common political belief system, rather than a direct influence. Hence, any approach for determining dependence needs to consider the possibility that the similarity between sources is due to the likelihood that the specific items are highly correlated.

**Incomplete Observations:** Our observations of data and updates may be incomplete for various reasons: some data sources may hide some of their data for security concerns; we may be unaware of some data sources that are commonly referred to for information; when the sources do not explicitly provide temporal information, the time values $time_j$ would need to be inferred from snapshots of the online data sources, taken only periodically or occasionally and possibly missing updates between consecutive observations. Incomplete observations can again lead to incorrect conclusions about the dependence between sources. Hence, any approach for determining dependence needs to consider the possibility that the granularity of observations introduces a degree of uncertainty about the information provided by the sources.

While the challenges identified above in determining dependence between sources are by no means exhaustive, they are certainly important issues, and any solution strategy would need to consider their impact.

## 3.2 Towards Solutions

We next discuss some preliminary ideas for developing robust solutions to identify dependence between data sources.

Generally speaking, dependence discovery can explore two underlying intuitions. First, in statistics, the probability of two independent events is the product of the probabilities of each individual event; any pair of events that violate this equation are dependent. We apply the same idea in source-dependence discovery. Consider two data sources $S_1$ and $S_2$ and we denote by $D_1$ and $D_2$ the data they provide respectively. If $Pr(D_1, D_2) \neq Pr(D_1) \cdot Pr(D_2)$, then $S_1$ and $S_2$ are likely to be dependent. This analysis requires us being able to compute the probability of a source providing a certain set of values.

The second intuition is that once we decide that two data sources are dependent, we consider the data source whose different subsets of data show different properties (*e.g.*, accuracy, average rating) as more likely to be dependent on the other. Formally, we denote by $F$ a property function of the data and divide $D_1$ into $D_1 \cap D_2$, the overlap between $S_1$ and $S_2$, and $D_1 - D_2$, the data that $S_1$ provides but $S_2$ does not. If $F(D_1 \cap D_2) \neq F(D_1 - D_2)$, $D_1$ is more likely to be dependent on $D_2$.

We next investigate two scenarios to show how these two underlying intuitions can be applied.

**Snapshot Dependence**

Consider the case where the information that is provided by a data source lacks temporal and probabilistic information, and we have only a single snapshot of each data source. Thus, each data source $S_i$ provides a single value $val_j$ for each identifier $id_j$, as in Table 1. A solution here needs to take two key intuitions into account:

1. Data sources that share common false values are much more likely to be dependent than data sources that share common true values, as the probability that two independent sources provide the same false value for a set of identifiers is typically low. This is akin to how teachers in schools determine if students may have copied from each other in a multiple-choice quiz. Note that this intuition permits accurate sources to have similar values, while still being considered as independent of each other.

**Table 3: Researcher affiliation example. Only $S_1$ provides up-to-date true values since 2002.**

|  | $S_1$ | $S_2$ | $S_3$ |
|---|---|---|---|
| Suciu | (2007, UW) (2006, MSR) (2002, UW) | (2006, MSR) (2001, UW) | (2003, UW) |
| Halevy | (2006, Google) (2002, UW) | (2006, Google) (2001, UW) | (2003, UW) |
| Balazinska | (2006, UW) | (2006, UW) | (2007, UW) |
| Dalvi | (2007, Yahoo!) (2002, UW) | (2007, Yahoo!) | (2003, UW) |
| Dong | (2007, AT&T) (2006, Google) (2002, UW) | (2006, Google) (2001, UW) | (2003, UW) |

2. If the accuracy of a data source (*e.g.*, fraction of true values) on the subset of information it shares in common with another data source is significantly different from its accuracy on the remaining information, the data source is more likely to be a partial copier than an independent data source. Note that this intuition is useful to identify copiers that copy only subsets of information from other sources.

To apply the above intuitions to determine dependence, it appears that one would require knowledge of what is true and what is false. If, however, one wishes to determine the truth from amongst conflicting values provided by multiple sources, knowledge of the dependence between sources is critical. A solution strategy can be devised using Bayesian analysis by iteratively determining true values, computing accuracy of sources, and discovering dependence between sources.

EXAMPLE 3.1. *Consider Example 2.1 and sources $S_1$–$S_5$ from Table 1. If we knew which values are true and which ones are false, we would suspect that $S_3$, $S_4$ and $S_5$ are dependent, because they provide the same false values. On the other hand, we would consider the dependence between $S_1$ and $S_2$ much less likely, as they share only true values. Based on this analysis, we could ignore the values provided by $S_4$ and $S_5$ during the voting process.* □

**Temporal Dependence**

Consider the case where the information provided by (or inferred from) a data source includes temporal information as well. Thus, each data source $S_i$ is associated with a set of $(time_j, val_j)$ pairs for each identifier $id_j$, as in Table 3. A solution here needs to reason over time, adding and refining to the key intuitions for snapshot dependence:

1. Data sources that share common false values are much more likely to be dependent than data sources that share common recent or outdated true values (*i.e.*, values that used to be true, but are no longer true). Note that outdated true values might have been characterized as false values, based on snapshot dependence.

2. Data sources that perform the same updates in close enough time frame are more likely to be dependent, especially if the same update trace is rarely observed from other sources. Note that this intuition prevents categorizing two sources that provide similar sets of identifiers as being dependent if their update traces are quite different.

3. If the accuracy of source $S_1$ on the subset of information it shares in common with, but provides earlier than, source

$S_2$ is significantly different from its accuracy on the subset of information it shares in common with, but provides later than, $S_2$, then $S_1$ and $S_2$ are likely to be similarity-dependent. Note that this intuition permits determination of dependence between sources even when the corresponding intuition for snapshot dependence may not suffice to determine dependence.

EXAMPLE 3.2. *Consider sources $S_1$, $S_2$ and $S_3$ in Table 3. Unlike in the case of Table 1, where sources $S_2$ and $S_3$ were assumed to provide false affiliations for some researchers, the availability of temporal information lets us infer that both $S_2$ and $S_3$ only provide out-of-date information, not false information.*

*The trace of updates makes us believe more that $S_2$ is independent of $S_1$, as many of its updates are before the corresponding ones by $S_1$, whereas $S_3$ is dependent on $S_1$, but just lazy in copying changes of values by $S_1$.* □

As in the case for snapshot dependence, a solution strategy can be devised based on an iterative scheme using Bayesian analysis, which would also need to address the following considerations:

- Discover dependence patterns of a data source over time. For example, a copier is more likely to remain as a copier; it can even choose to copy periodically and to copy from the same data sources.
- Distinguish between the update pattern of a copied value and that of an independently provided value. For example, if a source copies the value of a particular identifier, it may keep updating it from the original source. On the other hand, if a source provides the value of an identifier independently, it may not overwrite it in later copying.

## 4. APPLYING SOURCE DEPENDENCE

Knowledge of dependence between sources can improve a variety of technologies that have emerged over the years to help users access and digest information available from various sources. Applying such knowledge adds a new dimension to many existing data management tasks in data integration and data sharing, as discussed next.

**Data fusion:** Data fusion (surveyed in [3]) copes with combining conflicting data from multiple sources, but typically assumes independence of these sources. When deciding the truth from conflicting values, we would like to ignore values that are copied (but not necessarily the values independently provided by copiers). We can either determine one true value for each object, or identify a probabilistic distribution of possible values for each object and generate a probabilistic database.

When integrating answers from sources of probabilistic data, current techniques assume independence of sources and compute the probability of an answer tuple as the disjoint probability of its probabilities from each data source. Removing the independence assumption can significantly change the computation of the probabilities of the answer tuples.

**Record linkage:** Record linkage (surveyed in [17]) aims at linking representations of the same entities; however, in practice we often need to simultaneously conduct truth discovery and record linkage to distinguish between alternative representations and false values. Being able to link different representations of the same entity can often improve discovery of source dependence; on the other hand, knowing the dependence between data sources can help link different representations of the same entity. Thus, iterative strategies can simultaneously help in record linkage and in determining source dependence.

A challenge is that the boundary between a wrong value and an alternative representation is often vague. For example, "Luna Dong" is an alternative representation of "Xin Dong", while "Xing Dong" is a wrong value. How can one distinguish between them?

**Query answering:** When answering a (top-$k$) query, rather than necessarily going to all data sources and then combining the retrieved answers [10], we want to visit the most promising sources and avoid going to sources dependent on, or having been copied by, the ones already visited.

Challenges include formalizing the overlap of data sources based on dependence for query answering, adaptively deciding the next data source from which to retrieve data, and efficiently computing the probabilities and coverage percentages of current answers in an online fashion.

**Source recommendation:** An important issue in Web 2.0 applications or P2P applications is to identify sources or users that are trustable. Recommendations of such sources can be based on many factors, such as accuracy, coverage, freshness of provided data, and independence of opinions. While dependence in itself should be taken into consideration in source recommendation, it also implicitly affects our belief of other measures of a data source (through discovering truth and consensus opinions).

Note that whether we should recommend a dependent source is a tricky decision. If our goal is to find the truth or consensus opinion and avoid redundant information, we might prefer to ignore dependent sources; if our goal is to find diverse opinions, we might want to point out some sources that have dissimilarity-dependence on other sources.

We next illustrate the issues involved in some of these tasks using a real-life scenario.

EXAMPLE 4.1. *We extracted information on computer science books by searching* AbeBooks.com*, a listing-service website that integrates information on books from different online bookstores. In our collection there are 876 bookstores, 1263 books, and 24364 listings; each listing contains information including book title, author list, publisher, year, etc., on one book provided by one bookstore. A preliminary analysis of data from different bookstores reveals 471 pairs of bookstores that provide information on at least the same 10 books and are very likely to be dependent.*

*Now consider answering the following queries.*

1. *What are the books on Java Programming?*
2. *Who are authors of the book* Effective Java*?*
3. *Which books are authored by Jeffrey Ullman?*
4. *Who is the most productive publisher in the Database field?*

*Answering these queries accurately and efficiently places the following requirements on the system.*

1. *First, we would like to answer queries, such as Queries 2-4, accurately and completely. However, the data provided by the online bookstores are dirty for various reasons: the author lists are formatted in various ways; there are misspellings, missing authors, misordered authors, and wrong authors; extraction in itself can make mistakes (e.g., mistakes in parsing of authors, mis-interpretation of editors as authors), etc. Even after a pre-processing, the number of different author lists for each book varies from 1 to 23, and is 4 on average.*

   *To return quality answers, we need to be able to link different representations of values and resolve conflicting values.*

*One strategy that can be effective is to associate with each possible answer a probability indicating the likelihood that the answer is true given the conflicting information provided by different bookstores. The probability should capture our confidence of the extraction and data cleaning process, and take into consideration dependence between the bookstores.*

2. *Second, we would like to answer queries, such as Query 1 and Query 4, efficiently. (This requirement may not be prominent for this small data set, but is certainly necessary if we consider all books we can collect over the Web.) We observed that the quality of information provided by each bookstore varies: the number of computer science books provided by each bookstore varies from 1 to 1095, and the accuracy of the author lists provided by each bookstore, sampled on a set of 100 books, varies from 0 to .92.*

*To quickly retrieve query answers and reduce response time, we might adopt an online query answering approach, where we first return partially computed answers and then update probabilities of the answers as we query more data sources. In addition, we want to query the sources in an order such that we can return quality answers from the beginning. Identifying this order and computing the probabilities of answers require applying knowledge of dependence between sources and also accuracy of sources.* □

## 5. RELATED WORK

We are not aware of any existing work on detecting dependence between data sources and applying such dependence information in data integration, recommendation systems, and decision making. Recently there has been a lot of work studying how to manage *provenance* and *lineage* of data [4, 5, 8, 16]. These works all assume that the provenance or lineage information has already been provided by users or applications, and focus on how to effectively represent and retrieve such information. The scope of dependence we consider is broader: we examine not only copying of data, but also influence on opinions and conflicting data or opinions (dissimilarity-dependence). In addition, we study how to apply knowledge of dependence in the context of conflicting data to decide the true facts or the unbiased consensus of opinions.

The statistics community has studied the *opinion pooling* problem; that is, to combine probability distributions from multiple experts and arrive at a single probability distribution to represent the consensus behavior when the experts are dependent. Clemen and Winkler [7] showed that information from a set of dependent sources can be less valuable than that from independent sources and analyzed how shared information can lead to dependence between experts' opinions. In [6, 9, 13], Bayesian models are developed for combining probability distributions from dependent sources. However, these works did not study how to discover such dependence and the applications we consider face many other challenges, such as record linkage and query answering, in addition to combining various opinions.

Moss [15] studied how to detect plagiarism of programs by comparing *fingerprints* of the programs. We focus on structured information and analyze conflicting values or opinions provided by various sources. We expect that our techniques on dependence detection can also be adapted to extend plagiarism detection approaches.

Finally, the data cleaning community has studied how to use "dependency" information, such as functional dependencies and inclusion dependencies, to clean dirty data [14]. However, the dependencies they explore are different from source dependence that we consider in this paper.

## 6. CONCLUSIONS

The Web has accelerated the rate at which useful information is produced and disseminated, but has also eased the ability to spread false information. Whereas previous work on managing data from multiple sources focused on resolving heterogeneity of the data (including heterogeneity of the schema and of the representation of values), they often assumed consistency of the values and independence of sources. We argue that the next-generation data-sharing systems need to manage not only heterogeneity, but also conflicts and false information. In doing so, it is crucial that we are able to detect dependence between sources and leverage such information in a variety of technologies such as data integration and Web 2.0. In this paper, we have identified the types of dependence that are of particular interest in such systems, discussed preliminary ideas on how we can discover such dependence, and enumerated a few of the ways that we can apply the knowledge of dependence in managing data from multiple sources.

Considering dependence between data sources adds a new dimension to many existing data management topics, including but not limited to, data fusion, data cleaning, record linkage, distributed query answering, and recommendation systems. We have outlined several research opportunities that arise from taking a more principled view of source dependence. We expect research along this line can help users better understand data sources in the real world, and extract useful knowledge as they sail the ocean of information.